\documentclass[preprint]{aastex63}

\shorttitle{Detached Shell C Stars as TP Tracers}
\shortauthors{Kastner \& Wilson}

\begin{document}

\title{Detached Shell Carbon Stars: \\ Tracing Thermal Pulses on the Asymptotic Giant Branch}

\correspondingauthor{Joel Kastner}
\email{jhk@cis.rit.edu}

%\author[ORCID]{Joel H. Kastner}
\author{Joel H. Kastner}
\affiliation{Center for Imaging Science, 
  Rochester Institute of Technology, Rochester NY 14623, USA; jhk@cis.rit.edu}
\affiliation{School of Physics and Astronomy, 
  Rochester Institute of Technology} 
\affiliation{Laboratory for Multiwavelength Astrophysics, 
  Rochester Institute of Technology}

\author{Emily Wilson}
\affiliation{School of Physics and Astronomy, 
  Rochester Institute of Technology}
\affiliation{Center for Computational Relativity and Gravitation, 
  Rochester Institute of Technology}

%\tableofcontents

%\input{DSCstarPaperText}

\begin{abstract}
We consider whether the subset of carbon-rich asymptotic giant branch (AGB) stars that exhibit detached, expanding circumstellar shells may reveal the past histories of these stars as having undergone helium shell flashes (thermal pulses) on the AGB. We exploit newly available Gaia parallaxes and photometry, along with archival infrared photometry, to obtain refined estimates of the luminosities of all (12) known detached shell carbon stars. We examine the relationship between these luminosities and the estimated dynamical ages (ejection times) of the detached shells associated with the 12 stars, which range from $\sim$1000 to $\sim$30000 yr. 
%We find that carbon stars with the youngest detached shells (dynamical ages $<$3000 yr) have larger luminosities, on average, than those with intermediate-age detached shells (dynamical ages $\sim$8000--12000 yr), and that the luminosities of the carbon stars with the oldest shells (dynamical ages $\sim$20000--30000 yr) are similar to those of the C stars with young shells.
%This apparent overall decline in carbon star luminosity over a $\sim$$10^4$ yr timescale following shell ejection supports the hypothesis that the shell ejections are triggered by the sharp luminosity increases, and subsequent declines, that are predicted to define AGB thermal pulses. Furthermore, 
When arranged according to detached shell dynamical age, the (implied) luminosity evolution of the known detached shell carbon stars closely follows the predicted ``light curves'' of individual thermal pulses obtained from models of AGB stars. The comparison between data and models suggests that detached shell carbon stars are descended from $\sim$2.5--4.0 $M_\odot$ progenitors. We conclude that detached shell carbon stars may serve as effective tracers of the luminosity evolution of AGB thermal pulses.
\end{abstract}


\section{Introduction}

Intermediate-mass stars that undergo He shell burning on the asymptotic giant branch (AGB) may ultimately be responsible for the generation of most of the free carbon in the Universe \citep[see, e.g.,][and references therein]{Marigo2020}. If sufficient carbon is generated via nucleosynthesis in the deep AGB interior and mixed to the AGB stellar surface, carbon can become more abundant than oxygen by number, resulting in carbon star formation. The study of such carbon-rich AGB stars (hereafter C stars) is essential to trace the enrichment of the ISM in C, as well as in s-process elements that are generated via He shell burning \citep{Herwig2005}.

At solar metallicity, the onset of C/O $>$ 1 at an AGB star's surface  likely only occurs very late in AGB evolution, as a direct result of He shell flashes (thermal pulses; hereafter TPs) combined with repeated ``third dredge-up'' episodes, i.e., periods of internal convection and mixing that are sufficiently robust so as to bring the products of He shell burning to the surface \citep{Herwig2005}. Furthermore, inversion of the C/O ratio is only theoretically predicted to occur for a relatively narrow range of main-sequence progenitor masses, with the range being highly dependent on initial metallicity and the (poorly constrained) details of stellar mixing and convection \citep[e.g.,][]{Marigo2008}. In the Galaxy, there is observational evidence that only stars of initial mass $\ge$1.65 $M_\odot$ will evolve to become C stars \citep{Marigo2020}. The upper end of the C star progenitor mass range, which is more difficult to constrain observationally \citep[e.g.,][]{Kastner1993}, is predicted to lie at $\sim$4.5 $M_\odot$ for solar metallicity \citep{Karakas2014}. 

The C-rich dust generated in cool, highly extended C star photospheres has large opacity to optical/IR photons, making radiation pressure on dust a particularly efficient mass loss mechanism for such stars \citep{HofnerOlofsson2018}. As a consequence, most C stars exhibit large mass loss rates \citep[$\sim$$10^{-7}$--$10^{-5}$ $M_\odot$ yr$^{-1}$;][and references therein]{Olofsson1993, HofnerOlofsson2018}. Their expanding circumstellar molecular envelopes, readily detectable in millimeter-wave CO emission lines and the thermal infrared spectral signatures of C-rich dust, are characterized by large inferred CO:H$_2$ ratios and an array of trace C-bearing molecules \citep{HofnerOlofsson2018}. 

A subset of C stars also display extended, thin shells of ejecta that are detached from the stars. The first handful of such ``detached shell C stars'' were detected via analysis of their CO line profiles \citep{Olofsson1988,Olofsson1990,Olofsson1992}, with most of these detached shells later confirmed via single-dish and/or interferometric mapping \citep[e.g.,][]{Olofsson1996,Maercker2016, Kerschbaum2017}. Another handful, some with no known molecular gas (CO) counterparts, were detected via thermal IR imaging \citep{Izumiura1996,Kerschbaum2010,Mecina2014}. More recently, a detached shell associated with the C star TX Psc was identified via interferometric CO mapping \citep[][]{Brunner2019}. 

At the time of the discovery of the first such systems, it was hypothesized that each detached shell now coasting away from a C star is likely the direct result of the impulsive luminosity increase during a TP \citep{Olofsson1990,Olofsson1992}. Theoretical support for such a direct, causal connection has since steadily accumulated \citep[e.g.,][]{Steffen1998,Mattsson2007,Maercker2012}. In this paper, we explore one potential observational manifestation of the potential connection between C star TPs and detached shells: the relationship between the present-day luminosities of the dozen known detached shell C stars and the dynamical ages of their ejected shells. We exploit newly available Gaia parallaxes and photometry, along with archival infrared photometry, to calculate the luminosities of all (12) known detached shell carbon stars, and to obtain refined estimates of their shell ejection timescales. We place these results in the context of models of thermally pulsing AGB stars to assess whether detached shell carbon stars might trace the luminosity evolution of AGB thermal pulses.

\section{Sample and Data}\label{sec:Data}

The list of known detached shell C stars, compiled from \citet{Olofsson1988,Olofsson1990,Olofsson1992,Olofsson1996}, \citet{Izumiura1996}, \citet{Izumiura2011}, \citet{LindqvistOlofsson1999}, \citet{Schoier2005}, \citet{Kerschbaum2010}, \citet{Mecina2014}, and \citet{Brunner2019}, is presented in Table~\ref{tbl:PlxShell}. The Table lists parallaxes ($\pi$) for these stars as obtained from the Hipparcos and Gaia EDR3 catalogues \citep{vanLeeuwen2007,GaiaEDR3}. The Gaia EDR3 parallax uncertainties include the estimated systematic astrometric error ({\tt astrometric\_excess\_error}), added in quadrature to the formal parallax error. This is appropriate because the systematic error is likely to be statistically significant (i.e., {\tt astrometric\_excess\_noise\_sig} $>3$) for all of these stars \citep[see discussion in][their \S~5.3]{Lindegren2021}.
%The Table also lists detached shell angular radii ($\theta_s$) and expansion velocities ($v_s$), and the references for these shell parameters. 

\begin{table}
\begin{center}
\caption{\sc Detached Shell Carbon Stars: Parallaxes and Distances}
\label{tbl:PlxShell}
\footnotesize
\begin{tabular}{ccccccc}
\hline
Object & Hipparcos & \multicolumn{2}{c}{Gaia EDR3} & $1/\pi$ ($\sigma_{1/\pi}$) & Src.$^a$ & $D$ (BJ21$^b$) \\
           & $\pi$ (mas) & $\pi$ (mas) & RUWE & (pc) & & (pc) \\
\hline
\hline
U Cam &	0.79$\pm$1.35 & 1.62$\pm$0.29 & 0.95 &	618	(109) & G & 604$^{+12}_{-13}$ \\
DR Ser	& 0.59$\pm$1.48 & 0.91$\pm$0.22 & 1.00 & 	1103 (267) & G & 1054$^{+31}_{-32}$	\\
V644 Sco	&	... & 0.74$\pm$0.17 & 0.74 &	1353 (306)	& G & 1279$^{+50}_{-43}$ \\
R Scl & 2.11$\pm$1.75 &	2.54$\pm$0.56& 2.55 & 	394 (80) & G & 388$^{+12}_{-10}$ \\
TX Psc & 4.29$\pm$0.93 &	4.09$\pm$1.09& 1.92 &	233	(51) & H &	243$^{+12}_{-11}$ \\
U Ant	 &	3.9$\pm$0.67 &3.62$\pm$0.79& 1.03 &	256	(44) &	H & 276$^{+6}_{-7}$ \\
TT Cyg & 1.96$\pm$0.8 &	1.44$\pm$0.16& 1.03 &	695	(75) &	G & 618$^{+8}_{-10}$  \\
S Sct & 2.53$\pm$0.9 &	2.36$\pm$0.29 & 0.82 &	424	(53) &	G & 415$^{+10}_{-8}$ \\
U Hya & 6.18$\pm$0.75 &	... & ... &	162 (20) & H & ... \\
AQ And &	... &	1.26$\pm$0.15 & 1.00 &	794	(96)	& G & 769$^{+16}_{-18}$ \\
RT Cap &	1.78$\pm$1.48 &	1.80$\pm$0.24 & 0.82 &	556	(74) &	G & 542$^{+14}_{-13}$ \\
Y CVn &	4.59$\pm$0.73 &	3.22$\pm$1.30 & 1.97 &	218	(35) &	H & 314$^{+17}_{-18}$ \\
\hline
\end{tabular}
\end{center}

{\sc Notes:} 
a) Source for distance in column 5: G = straight inverse of Gaia EDR3
parallax; H = straight inverse of Hipparcos parallax. b) Gaia EDR3
distance from \citet{BailerJones2021}. 
\end{table}

\begin{table}
\begin{center}
\caption{\sc Detached Shell Carbon Stars: Photometry$^a$}
\label{tbl:Phot}
\tiny
\begin{tabular}{ccccccccccccccccccccc}
\hline
Band$^b$	&	$B_P$	&	$V$	& $G$	&	$R_P$	&	$J$	&	$H$	&	$K_s$	&	W1	&	W2	&	S9W	&	W3	&	12	&	L18W	&	W4	&	25	&	60	&	100	&	N160 & $E(B-V)$ 	\\
%		&	freq	&	595300	&	541430	&	514900	&	393400	&	241960	&	181750	&	138550	&		&		&	34819	&		&	25866	&	16302	&		&	12554	&	4847.1	&	2940.6	&		\\
$\lambda$ ($\mu$m)	&	0.50	&	0.55	&	0.58	&	0.76	&	1.24	&	1.65	&	2.17	&	3.35	&	4.60	&	8.62	&	11.57	&	11.60	&	18.4	&	22.1	&	23.9	&	61.9	&	102	&	160	\\
\hline
\hline
	U Cam	&	3.14	&	3.51	&	16.4	&	41.1	&	187	&	395	&	422	&		&		&	133	&		&	121	&	39.3	&		&	40.9	&	16.9	&	7.31	&	& 0.13 	\\
%		&		&	595300	&	541430	&	514900	&	393400	&	241960	&	181750	&	138550	&	89490	&	65172	&	34819	&	25934	&	25866	&	16302	&	13571	&	12554	&	4847.1	&		&		\\
%		&		&	0.50	&	0.55	&	0.58	&	0.76	&	1.24	&	1.65	&	2.17	&	3.35	&	4.60	&	8.62	&	11.57	&	11.60	&	18.40	&	22.11	&	23.90	&	61.89	&		&		\\
	DR Ser	&	0.423	&	0.65	&	2.58	&	6.69	&	46.2	&	106	&	124	&	47.5	&	25.7	&	21	&	11	&	15.9	&	6.16	&	5.56	&	6.66	&	5.96	&		& & 0.58		\\
%		&		&	595300	&	541430	&	514900	&	393400	&	241960	&	181750	&	138550	&		&		&	34819	&		&	25866	&	16302	&		&	12554	&	4847.1	&	2940.6	&	1873.7	\\
%		&		&	0.50	&	0.55	&	0.58	&	0.76	&	1.24	&	1.65	&	2.17	&		&		&	8.62	&		&	11.60	&	18.40	&		&	23.90	&	61.89	&	102.02	&	160.11	\\
	V644 Sco	&	0.386	&	0.28	&	2.44	&	6.29	&	38.3	&	88.1	&	105	&	39.2	&	35	&	22.7	&	10.8	&	17.6	&		&	8.26	&	8.99	&		&		& & 0.26		\\
%		&		&	595300	&	541430	&	514900	&	393400	&	241960	&	181750	&	138550	&	85118	&	61312	&	34819	&	25934	&	25866	&	16302	&	13571	&	12554	&	4847.1	&	2940.6	&		\\
%		&		&	0.50	&	0.55	&	0.58	&	0.76	&	1.24	&	1.65	&	2.17	&	3.52	&	4.89	&	8.62	&	11.57	&	11.60	&	18.40	&	22.11	&	23.90	&	61.89	&	102.02	&		\\
	R Scl	&	9.03	&	8.57	&	38.2	&	92.2	&	256	&	554	&	752	&		&		&	194	&		&	162	&	65.3	&		&	82.1	&	54.8	&	23.2	&	5.59 & 0.01	\\
%		&		&	595300	&	541430	&	514900	&	393400	&	241960	&	181750	&	138550	&		&		&	34819	&	25934	&	25866	&	16302	&	13571	&	12554	&	4847.1	&	2940.6	&		\\
%		&		&	0.50	&	0.55	&	0.58	&	0.76	&	1.24	&	1.65	&	2.17	&		&		&	8.62	&	11.57	&	11.60	&	18.40	&	22.11	&	23.90	&	61.89	&	102.02	&		\\
	TX Psc	&	25.5	&	35.1	&	92.1	&	192	&	524	&	1060	&	1080	&		&		&	174	&	95.7	&	150	&	45.4	&	32.6	&	41.9	&	13.4	&	7.27	&	& 0.08	\\
%		&		&	595300	&	541430	&	514900	&	393400	&	241960	&	181750	&	138550	&	89490	&	65172	&	34819	&	25934	&	25866	&		&	13571	&	12554	&		&		&		\\
%		&		&	0.50	&	0.55	&	0.58	&	0.76	&	1.24	&	1.65	&	2.17	&	3.35	&	4.60	&	8.62	&	11.57	&	11.60	&		&	22.11	&	23.90	&		&		&		\\
	U Ant	&	16.7	&	23	&	68.2	&	158	&	591	&	1190	&	1270	&	717$^c$	&	224$^c$	&	264	&	111	&	168	&	61.5	&	35.8	&	44.8	&	27.1	&	21.1	&	& 0.04	\\
%		&		&	595300	&	541430	&	514900	&	393400	&	241960	&	181750	&	138550	&		&		&	34819	&	25934	&	25866	&	16302	&	13571	&	12554	&	4847.1	&		&		\\
%		&		&	0.50	&	0.55	&	0.58	&	0.76	&	1.24	&	1.65	&	2.17	&		&		&	8.62	&	11.57	&	11.60	&	18.40	&	22.11	&	23.90	&	61.89	&		&		\\
	TT Cyg	&	2.69	&	3.23	&	9.19	&	20.6	&	52.1	&	102	&	111	&		&	37.6	&	20.4	&	8.08	&	15.8	&	5.37	&	3.37	&	4.17	&	3.45	&	4.41	& & 0.09		\\
%		&		&	595300	&	541430	&	514900	&	393400	&	241960	&	181750	&	138550	&		&		&	34819	&	25934	&	25866	&	16302	&	13571	&	12554	&	4847.1	&	2940.6	&	1873.7	\\
%		&		&	0.50	&	0.55	&	0.58	&	0.76	&	1.24	&	1.65	&	2.17	&		&		&	8.62	&	11.57	&	11.60	&	18.40	&	22.11	&	23.90	&	61.89	&	102.02	&	160.11	\\
	S Sct	&	5.18	&	6.81	&	21.5	&	50.7	&	189	&	368	&	379	&		&		&	78.7	&	50.2	&	65.3	&	20.1	&	13.2	&	17.3	&	9.28	&		&	& 0.10	\\
%		&		&	595300	&	541430	&	514900	&	393400	&	241960	&	181750	&	138550	&		&	65172	&	34819	&	25934	&	25866	&	16302	&	13571	&	12554	&	4847.1	&	2940.6	&		\\
%		&		&	0.50	&	0.55	&	0.58	&	0.76	&	1.24	&	1.65	&	2.17	&		&	4.60	&	8.62	&	11.57	&	11.60	&	18.40	&	22.11	&	23.90	&	61.89	&	102.02	&		\\
	U Hya	&	27.8	&	40.3	&	92.5	&	194	&	753	&	1330	&	1310	&		&		&	257	&	87.1	&	206	&	78.1	&	56.5	&	72.4	&	17.2	&	14.5	&	0.766 & 0.00	\\
%		&		&	595300	&	541430	&	514900	&	393400	&	241960	&	181750	&	138550	&		&		&	34819	&	25934	&	25866	&	16302	&	13571	&	12554	&	4847.1	&	2940.6	&		\\
%		&		&	0.50	&	0.55	&	0.58	&	0.76	&	1.24	&	1.65	&	2.17	&		&		&	8.62	&	11.57	&	11.60	&	18.40	&	22.11	&	23.90	&	61.89	&	102.02	&		\\
	AQ And	&	1.75	&	3.11	&	7.42	&	17.2	&	54.1	&	119	&	143	&		&		&	32	&	19.7	&	25.7	&	8.16	&	6.38	&	7.41	&	3.83	&	4.4	&	& 0.07	\\
%		&		&	595300	&	541430	&	514900	&	393400	&	241960	&	181750	&	138550	&	89490	&	65172	&	34819	&	25934	&	25866	&	16302	&	13571	&	12554	&	4847.1	&	2940.6	&		\\
%		&		&	0.50	&	0.55	&	0.58	&	0.76	&	1.24	&	1.65	&	2.17	&	3.35	&	4.60	&	8.62	&	11.57	&	11.60	&	18.40	&	22.11	&	23.90	&	61.89	&	102.02	&		\\
%		&		&	595300	&	541430	&	514900	&	393400	&	241960	&	181750	&	138550	&		&		&	34819	&	25934	&	25866	&	16302	&	13571	&	12554	&	4847.1	&	2940.6	&	1873.7	\\
%		&		&	0.50	&	0.55	&	0.58	&	0.76	&	1.24	&	1.65	&	2.17	&		&		&	8.62	&	11.57	&	11.60	&	18.40	&	22.11	&	23.90	&	61.89	&	102.02	&	160.11	\\
	RT Cap	&	4.09	&	5.93	&	18.2	&	44.6	&	231	&	449	&	510	&		&		&	86.6	&	43.5	&	72.9	&	21.7	&	15.9	&	20.7	&	4.41	&	3.61	&	3.79	& 0.13 \\
	Y CVn	&	23.1	&	30	&	84.3	&	181	&	634	&	1360	&	1330	&		&		&	304	&	193	&	276	&	91.7	&	57	&	70.3	&	17.2	&	7.82	&	& 0.03	\\
\hline
\end{tabular}
\end{center}

{\sc Notes:}
a) Fluxes in Jy. b) $B_PGR_p$ from Gaia EDR3; $JHK_s$ from 2MASS; W1--4 from WISE; 12, 25, 60, 100 from IRAS; S9W, L18W, N160 from AKARI. c) 3.52 $\mu$m and 4.89 $\mu$m fluxes from DIRBE.
\end{table}

The general utility of Gaia parallax measurements for AGB stars, at least for the more nearby examples, is questionable, due to their large fluxes in the red combined with the fact that their spatially variable photospheres can be resolved by Gaia \citep{VanLangevelde2018,Vlemmings2019,Ramstedt2020}. However, Table~\ref{tbl:PlxShell} demonstrates that the Gaia EDR3 and Hipparcos parallaxes are in agreement, to within the errors, for all stars for which both data sources are available --- including the four stars for which both astrometry missions yield parallax measurements with signal/noise ratio $>$2.5 (TX Psc, U Ant, S Sct, Y CVn), as well as several stars for which the Hipparcos measurements are of lower significance (R Scl, TT Cyg, RT Cap). This comparison between Gaia EDR3 and Hipparcos parallaxes in Table~\ref{tbl:PlxShell} indicates that the former are, overall, viable measurements for these stars, notwithstanding the fact that three stars have renormalized unit weight error (RUWE) values $>1$ (see Table~\ref{tbl:PlxShell}), which is a potential indicator of a poor astrometric fit \citep[][]{Lindegren2021}. Indeed, we note that the Gaia EDR3 parallax measured for R Scl --- the sample star with the largest RUWE --- is in excellent agreement with the distance independently obtained from analysis of light echoes \citep[361$\pm$44 pc;][]{Maercker2018b}. 

We have thus employed Gaia EDR3-based distances for the analysis described in this paper. Specifically, we have adopted the Gaia EDR3-based distances to these stars obtained by \citet{BailerJones2021} from their Bayesian statistical approach, which also accounts for the parallax zero-point offset \citep[although this correction is likely to be negligibly small for the sample stars;][]{Lindegren2021}. In Table~\ref{tbl:PlxShell}, we list these distances ($D$), alongside the distance to each star (and associated distance uncertainty) obtained as the straight inverse of the Gaia or Hipparcos parallax, where we only list the inverse parallax with the greater significance. It is evident that the two distance estimates are in good agreement, with the \citet{BailerJones2021} distances systematically smaller and more (formally) precise than the straight inverse of the Gaia parallaxes, as expected \citep[see discussion in ][]{Bailer-Jones2015}. The sole exception is Y CVn, for which the inverse of the Hipparcos parallax is significantly smaller than the  \citet{BailerJones2021} Gaia distance. This case is discussed further in \S~\ref{sec:LbolShells}.

%; we further describe and support this choice of distances in \S~\ref{sec:LbolShells}.

Photometric data from the Gaia EDR3, 2MASS, WISE, AKARI, and/or IRAS archives, obtained via the Vizier photometry tool\footnote{{\tt http://vizier.u-strasbg.fr/vizier/sed/doc/}}, are listed in Table~\ref{tbl:Phot}. These data were used to compile spectral energy distributions (SEDs; Fig.~\ref{fig:SEDs}) for purposes of  determining bolometric fluxes  en route to calculating bolometric luminosities (\S~\ref{sec:LbolShells}). 
The Gaia EDR3 ($B_PGR_P$) and 2MASS ($JHK_s$) photometric data were dereddened to account for the estimated line-of-sight interstellar extinction toward each star ($E(B-V)$, listed in Table~\ref{tbl:Phot}, last column) as obtained from the {\tt Bayestar19} \citep[][]{Green2019} or {\tt Stilism} \citep[][]{Lallement2019} Galactic dust extinction maps\footnote{{\tt Bayestar19: http://argonaut.skymaps.info/; Stilism: https://stilism.obspm.fr/}}. For stars with reddening measurements available in both maps (e.g., DR Ser), we report the {\tt Bayestar19} value of $E(B-V)$; however, we note that the two maps yield consistent $E(B-V)$ estimates (in the case of DR Ser, which suffers the largest reddening, the Stilism-based estimate of $E(B-V) = 0.56$ is in excellent agreement with the Bayestar19-derived value listed in Table 2). Photometric corrections based on $E(B-V)$ were then applied according to the relations in \citet{Schlafly2011}.  

\section{Detached Shell C Stars: Luminosities and Shell Ages}\label{sec:LbolShells}

\begin{table}
\begin{center}
\caption{\sc Detached Shell Carbon Stars: Luminosities and
  Shell Parameters}
\label{tbl:DistLumShells}
\footnotesize
\begin{tabular}{ccccccccc}
\hline
Object & $D^a$ & $L_{\rm bol}$$^b$ & $L_{\rm bol}$ (Ref.)$^c$ & $\theta_s$ &  $v_s$ & Refs.$^d$ &  $R_s$ &  $\tau_s$  \\
           &  (pc) &  ($L_\odot$)  &   ($L_\odot$)  & ($''$) & (km s$^{-1}$) & & ($10^4$ au) & (yr) \\
\hline
\hline
U Cam & 604 & 	11800 & 13900 (1) & 7.3 &	23  & 1 &	0.44 &	930 \\
DR Ser & 1054 & 12400 & 7700 (1) & 7.6 &	20 & 2, 3 & 0.80 & 1940 \\
V644 Sco & 1279 & 11300 & 15000 (1) & 9.4 &	23.2 & 4, 5 &	1.2 &	2530 \\
R Scl & 388 &	6750 & 7700 (1) & 	19.5 &	14.3 & 6 & 0.76 &	2560 \\
TX Psc & 243 &	5300 & 6000 (2) & 22 &	10  & 7 & 0.54 &	2590 \\
U Ant & 276 & 	7100 & 6060 (1) & 42.5	& 19  & 8 & 1.2 & 3000 \\
TT Cyg & 618 &	3350 & 3300 (1) & 35 &	13.3 & 5& 2.2 & 8100 \\
S Sct & 415 &	5200 & 5300 (1) & 70 & 	17.3 & 5 & 2.9 & 8300 \\
U Hya & 162 & 2800 & 2960 (3) & 110 & 7.6$^e$ & 9 & 1.8 & 12000 \\
AQ And & 769 & 6150 & 10400 (4) & 55	&	10$^f$ & 10 &	4.2 & 21000  \\
RT Cap & 542 & 10900 & 8500 (5) & 94 &	8$^e$ & 11, 12 &	 5.1 &	31000  \\
Y CVn & 314 & 10600 & 9100 (5) 	& 190 &	8$^e$ & 13, 12  & 6.0 &	36000 \\
\hline
\end{tabular}
\end{center}

{\sc Notes:} \\
a) From Table \ref{tbl:PlxShell}.\\
b) $L_{\rm bol}$ obtained from our blackbody fitting (see \S
\ref{sec:LbolShells}). 
c) Literature $L_{\rm bol}$ value (with reference in parentheses),
corrected for the distance listed in column 2. \\
References: 1. \citet{Schoier2005}; 2. \citet{Brunner2019};
3. \citet{Izumiura2011}; 4. \citet{Kerschbaum2010};
5. \citet{Schoier2001}. \\
d) References for detached shell angular diameters ($\theta_s$) and
expansion velocities ($v_s$): 
1. \citet{LindqvistOlofsson1999};
2. \citet{Ramstedt2011}; 
3. \citet{Schoier2005};
4. \citet{Maercker2018b};
5. \citet{Olofsson1996}; 
6. \citet{Maercker2016}; 
7. \citet{Brunner2019};
8. \citet{Kerschbaum2017};
9. \citet{Izumiura2011}; 
10. \citet{Kerschbaum2010}; 
11. \citet{Mecina2014}. 
12. \citet{Olofsson1993};
13. \citet{Izumiura1996}.\\
e) Expansion velocity obtained from spatially
integrated CO line profile; possibly dominated by present-day mass
loss. \\
f) Assumed shell expansion velocity. 
\end{table}

% 0 name, x, xerr, y, yerr: UCam 950.0 167.5566343042071 11700 -3763.2175511358278
% 1 name, x, xerr, y, yerr: DRSer 2030.0 491.3961922030825 7040 -2995.785433117789
% 2 name, x, xerr, y, yerr: RScl 2380.0 290.0831024930748 5770 -1320.8204356934023
% 3 name, x, xerr, y, yerr: TXPsc 2480.0 542.8326180257511 4570 -1781.6509790196908
% 4 name, x, xerr, y, yerr: V644Sco 2670.0 603.8580931263858 8840 -3546.4132428060834
% 5 name, x, xerr, y, yerr: UAnt 2780.0 477.8125 4840 -1520.771484375
% 6 name, x, xerr, y, yerr: SSct 8450.0 1056.25 4620 -1082.8125
% 7 name, x, xerr, y, yerr: TTCyg 9050.0 976.6187050359712 3590 -733.0133015889446
% 8 name, x, xerr, y, yerr: UHya 12300.0 1518.5185185185185 2540 -588.4468830970887
% 9 name, x, xerr, y, yerr: AQAnd 21100.0 2551.133501259446 5960 -1354.0830790119858
% 10 name, x, xerr, y, yerr: YCVn 25000.0 4013.7614678899085 4550 -1343.7263277501895
% 11 name, x, xerr, y, yerr: RTCap 31600.0 4205.755395683454 9700 -2410.189689974638

%%% us vs Schoier2005:
% R Scl: 5770 vs 6660
% U Cam: 11700 vs 14500
% U Ant: 4840 vs 5620
% V644 Sco: 8840 vs 15000
% DR Ser: 7047 vs 8400
% S Sct: 4620 vs 5500
% TT Cyg: 3590 vs 4150
%%% us vs Brunner+19:
% TX Psc: 4570 vs 5530
%%% us vs  Izumiura+11 who say ..."a temperature of 2800 K with a luminosity of 2960 L located at 162 pc from the Sun"
% U Hya: 4190 vs 2960
%%% us vs Kerschbaum2010}:
% AQ And: 5960 vs 11100 (from P-L relation)
%%% us vs Schoier2001:
% Y CVn: 4550 vs 4400
% RT Cap: 9700 vs 9000

\begin{figure}[!ht]
\begin{center}
\includegraphics[width=7in]{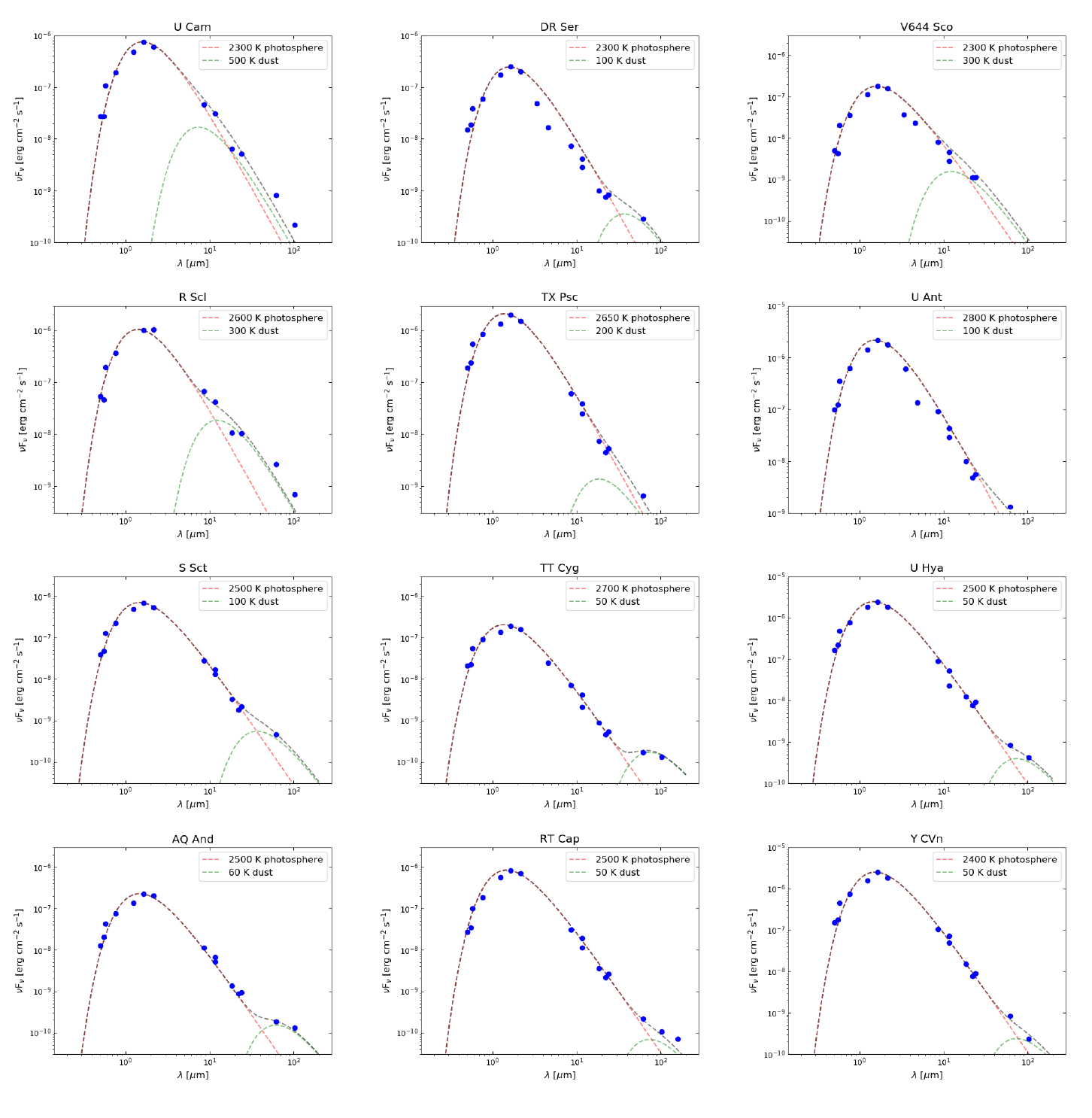}
\end{center}
\caption{Optical through IR photometry for carbon stars with detached shells. Data are derived from the Gaia, 2MASS, WISE, AKARI, and/or IRAS archives (Table~\ref{tbl:Phot}). The Gaia and 2MASS photometric data have been dereddened as described in the text. A pair of blackbodies approximating the contributions of stellar photosphere (red dashed curve) and cooler circumstellar dust (green dashed curve) is overlaid on each star's photometry; the black dashed curve represents the composite model.}
\label{fig:SEDs}
\end{figure}

To obtain luminosities for the detached shell C stars, we have calculated their bolometric fluxes ($F_{\rm bol}$) from the data listed in Table~\ref{tbl:Phot}. To obtain $F_{\rm bol}$ for each star, we fit a composite of two blackbodies to its dereddened SED, as an approximation of the combination of stellar photosphere plus a contribution from cool circumstellar dust; the photospheric blackbody was normalized to each star's H band flux. The blackbody fits (Fig.~\ref{fig:SEDs}) demonstrate that, in all cases, the stellar photospheric emission dominates the bolometric flux \citep[consistent with the results of][]{Schoier2005}. %The resulting values of $F_{\rm bol}$ and $T_{\rm eff}$ are listed in Table~\ref{tbl:PlxFluxesTeff}. 
The best-fit blackbody-equivalent effective temperatures ($T_{\rm eff}$) we find for the detached shell C stars (2300--2800 K; Fig.~\ref{fig:SEDs}) are generally consistent with previous blackbody-fit-based estimates for many of these same stars \citep[e.g.,][]{Schoier2001,Maercker2018b}. These temperatures are systematically lower than the $T_{\rm eff}$ values obtained by \citet{Bergeat2001} via modeling of C star atmospheric spectral features, for the stars in common. However, we find that the Wien tails of blackbodies fixed to the \citet{Bergeat2001} $T_{\rm eff}$ values overshoot the stars' optical fluxes, with the result that $F_{\rm bol}$ is overestimated, typically by $\sim$10--15\%. Thus, in all cases, our calculations of $L_{\rm bol}$ are based on the integrated fluxes of the best-fit blackbodies plotted in Fig.~\ref{fig:SEDs}.

The resulting estimates of $L_{\rm bol}$, as obtained from the blackbody fits and the distances listed in column 2 of Table~\ref{tbl:DistLumShells}, are listed in Table~\ref{tbl:DistLumShells}, column 3. The Table also lists $L_{\rm bol}$ values for the same stars as gleaned from the literature and (for consistency) corrected for the Gaia EDR3-based distances listed in the Table. All but two of our luminosity estimates are within $\sim$30\% of the literature values. The two exceptions are DR Ser, for which the {\tt Bayestar19} map yields an estimate of ISM extinction that is a factor $\sim$2 larger than previously assumed \citep{Schoier2005}, and AQ And, for which the literature estimate was based on a period-luminosity relationship rather than integrated flux \citep{Kerschbaum2010}.
%Thus, while the individual $L_{\rm bol}$ values typically have $\sim$30\% uncertainties, the {\it relative} luminosities of the detached shell C stars listed in Table~\ref{tbl:DistLumShells} --- both as derived here, and as previously derived by others --- would appear to be well constrained. 
Henceforth, we adopt the $L_{\rm bol}$ values obtained from our blackbody fits to dereddened photometry (i.e., the values listed in the third column of Table~\ref{tbl:DistLumShells}).

Table~\ref{tbl:DistLumShells} also lists the dynamical age of the detached shell associated with each C star, $\tau_s$. These values are obtained from the shell's angular radius ($\theta_s$) as reported in CO mapping observations and/or thermal infrared emission from dust; the adopted distance, which yields the shell linear radius $R_s$; and the shell expansion speed ($v_s$) as measured in mm-wave (rotational) lines of CO. The values of $\theta_s$ and $v_s$ adopted for these calculations (and the references for these parameters), along with the estimates for $R_s$, are also compiled in Table~\ref{tbl:DistLumShells}. For the handful of cases where CO emission has yet to be detected from the shells themselves --- three of which have the largest linear radii --- we have either adopted the expansion velocity measured for the present-day CO mass loss envelope or (in one case) assumed an expansion speed of 10 km s$^{-1}$ (see Table~\ref{tbl:PlxShell}). The dynamical ages for these shells in particular --- and for the larger and (hence) more evolved shells more generally --- may be somewhat overestimated, if the shells had higher initial $v_s$ (i.e., similar to the speeds of the smaller detached shells) and then were significantly decelerated as they encountered and swept up mass previously ejected by the central AGB stars \citep[][]{Steffen1998,Mattsson2007,Libert2007,Matthews2013}. A deeper spectroscopic search for molecular or atomic gas in these very extended shells is clearly warranted, so as to establish the shell expansion speeds \citep[indeed, deep H {\sc i} observations indicate the present-day expansion speed of the Y CVn detached shell is only $\sim$1--2 km s$^{-1}$;][]{Libert2007}. It is possible that any CO in these old and very extended shells has been dissociated by interstellar UV, such that a search for 492 GHz C {\sc i} line emission might be more fruitful \citep[see][]{Olofsson2015}, although this line may not be efficiently excited at the expected low temperatures and densities that should characterize these shells.

\begin{figure}[!ht]
\begin{center}
\includegraphics[width=6in]{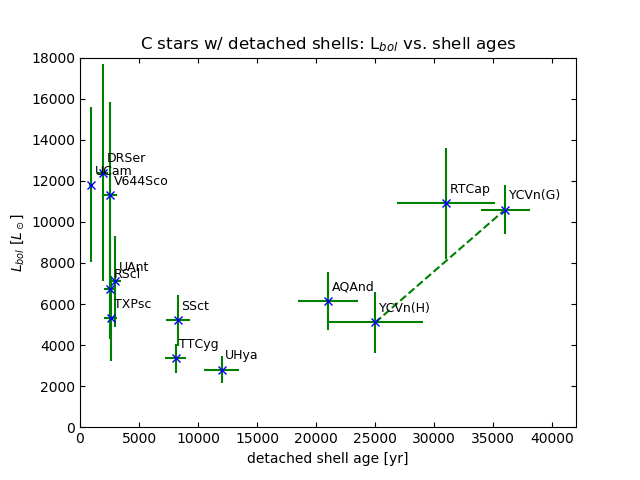}
\end{center}
\caption{Bolometric luminosity ($L_{\rm bol}$) vs.\ shell dynamical age ($\tau_s$) for detached shell C stars (see Table~\ref{tbl:DistLumShells} and \S~\ref{sec:LbolShells}). The two points plotted for Y CVn represent its ($L_{\rm bol}$, $\tau_s$) positions assuming its Hipparcos (H) vs.\ \citet{BailerJones2021} Gaia EDR3 (G) distances.}
\label{fig:LbolVsAge}
\end{figure}

A plot of $L_{\rm bol}$ vs.\ $\tau_s$, obtained from the parameter values in Table~\ref{tbl:DistLumShells}, is presented in Fig.~\ref{fig:LbolVsAge}. To estimate the uncertainties on the two quantities, we have (somewhat pessimistically) assumed that the (relative) parallax errors, which are typically $\sim$15--20\%, dominate the error propagation. This results in uncertainties in $L_{\rm bol}$ ranging from $\sim$20\% to $\sim$40\%. If we were to instead adopt the \citet{BailerJones2021} distance uncertainties (which are typically a few percent; Table~\ref{tbl:DistLumShells}), the ($\sim$10--30\%) uncertainties that enter into the calculation of $F_{\rm bol}$ (see above) would then become the important or perhaps dominant source of error in $L_{\rm bol}$, while the uncertainties in $\tau_s$ would simply shrink in direct proportional to the uncertainties in $D$. We further note that (as was just implied) the $L_{\rm bol}$ and $\tau_s$ uncertainties are correlated, since both quantities are dependent on the adopted distance. The star Y CVn well illustrates this effect: in Fig.~\ref{fig:LbolVsAge}, we have plotted its ($L_{\rm bol}$, $\tau_s$) positions as obtained from its \citet{BailerJones2021} distance and as obtained from the inverse of its Hipparcos parallax. The discrepancy in distance translates to a significant diagonal translation of the star in $L_{\rm bol}$ vs.\ $\tau_s$ space. However, as Y CVn is the only star for which the \citet{BailerJones2021} and inverse (Hipparcos) parallax distances are discrepant, it represents an extreme (indeed, unique) case in this regard.

Notwithstanding the foregoing caveats, Fig.~\ref{fig:LbolVsAge} appears to reveal a U-shaped dependence of C star luminosity on detached shell age: the three C stars with the youngest detached shells (U Cam, DR Ser, V644 Sco) have the largest luminosities ($\sim$$10^4$ $L_\odot$); the next three stars in detached shell age order (R Scl, TX Psc, U Ant) have somewhat ($\sim$40\%) smaller luminosities; the next three stars (TT Cyg, S Sct, U Hya), which have ``intermediate-age'' shells (dynamical ages 8000--12000 yr), have the smallest luminosities (a few $\times$$10^3$ $L_\odot$); and, finally, the three stars (AQ And, RT Cap, Y CVn) with the oldest shells (ages $\sim$20000-35000 yr) appear to ``rebound'' to luminosities in the range $\sim$5000--10000 $L_\odot$. Fig.~\ref{fig:LbolVsAge} thus suggests that detached shell C stars may be undergoing significant luminosity evolution on shell ejection timescales. Furthermore,
the collective $L_{\rm bol}$ vs.\ $\tau_s$ behavior apparent in Fig.~\ref{fig:LbolVsAge} --- an initial steep decline in $L_{\rm bol}$, followed by a slow recovery --- is reminiscent of the time evolution of an individual AGB thermal pulse, as predicted by theory \citep[see, e.g., Fig.~1 in][]{Mattsson2007}. 

%This apparent relationship between the luminosities of detached shell C stars and their shell dynamical ages motivates the following exploration of AGB star thermal pulse models.
%Indeed, from Table~\ref{tbl:DistLumShells}, we find that this difference is likely to be statistically significant. The weighted means of the (six) present-day bolometric luminosities of the carbon stars with young shells vs.\ the (six) stars with intermediate-age to ``old'' shells (as just defined) are  $\overline{L} = 7500\pm1200$ $L_\odot$ vs.\ $\overline{L} = 4200\pm500$ $L_\odot$, respectively; the likelihood that these two samples have the same mean luminosity is only 5.4\%. 
%As noted earlier, the relative luminosities of the detached shell C stars are well constrained, so the foregoing difference in mean stellar luminosities for stars with younger vs.\ older shells appears to be a robust result. Thus, the $L_{\rm bol}$ vs.\ $\tau_s$ data for detached shell C stars appear to reveal an overall decline in the luminosities of these stars over the few thousand years following their shell ejections. 

\section{Comparison with Models}

To further investigate the potential relationship between $L_{\rm bol}$ and $\tau_s$ for these stars, we generated models of the temporal luminosity behavior of thermally pulsing AGB stars. We used Modules for Experiments in Stellar Astrophysics \citep[\textsc{MESA};][]{Paxton2011,Paxton2018}, release 10108, to simulate the temporal luminosity evolution of stars with initial masses 2.0, 2.8, and 3.4 M$_\odot$ from the zero-age main sequence through AGB stages. The models considered here hence span a range of initial masses appropriate for C star generation on the AGB, and span the stars' complete post-main sequence evolution at time resolutions sufficient to trace individual TPs. In the resulting MESA models, the AGB phase is but a few percent of the MS lifetime, consistent with previous detailed modeling \citep[e.g.,][]{Herwig2005}. 

\begin{figure}[!ht]
\begin{center}
\includegraphics[width=6in]{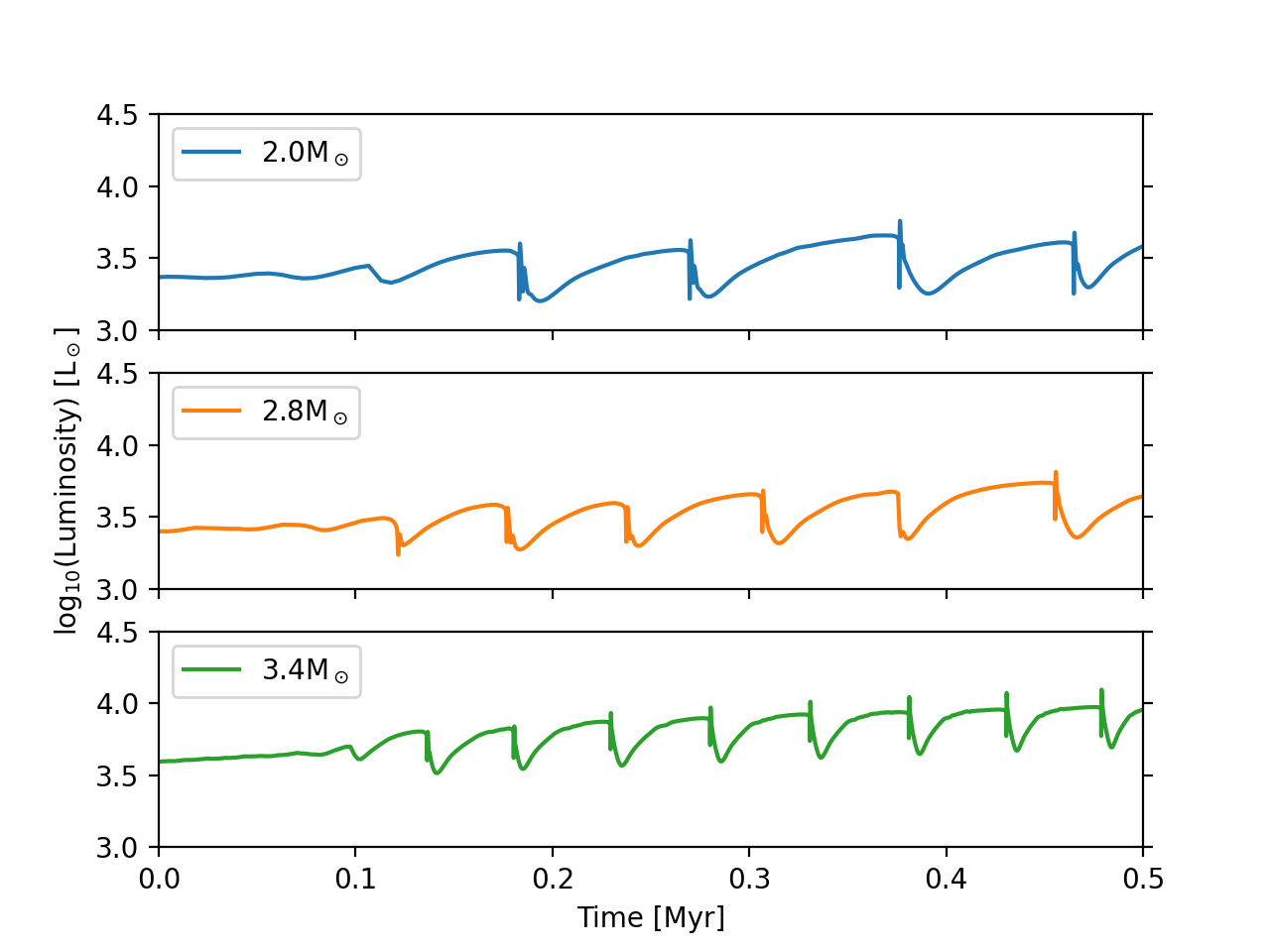}
\end{center}
\caption{The late AGB portions of the 2.0, 2.8, and 3.4 M$_\odot$ MESA models. The three time series shown here span $5\times10^5$ yr and have been aligned by applying ``zero-point'' time offsets of $\sim$$1.2\times10^9$, $5.7\times10^8$, and $3.1\times10^8$ yr, respectively, such that the onset of their TP stages roughly coincide. }
\label{fig:MESA}
\end{figure}

\begin{figure}[!ht]
\begin{center}
\includegraphics[width=6in]{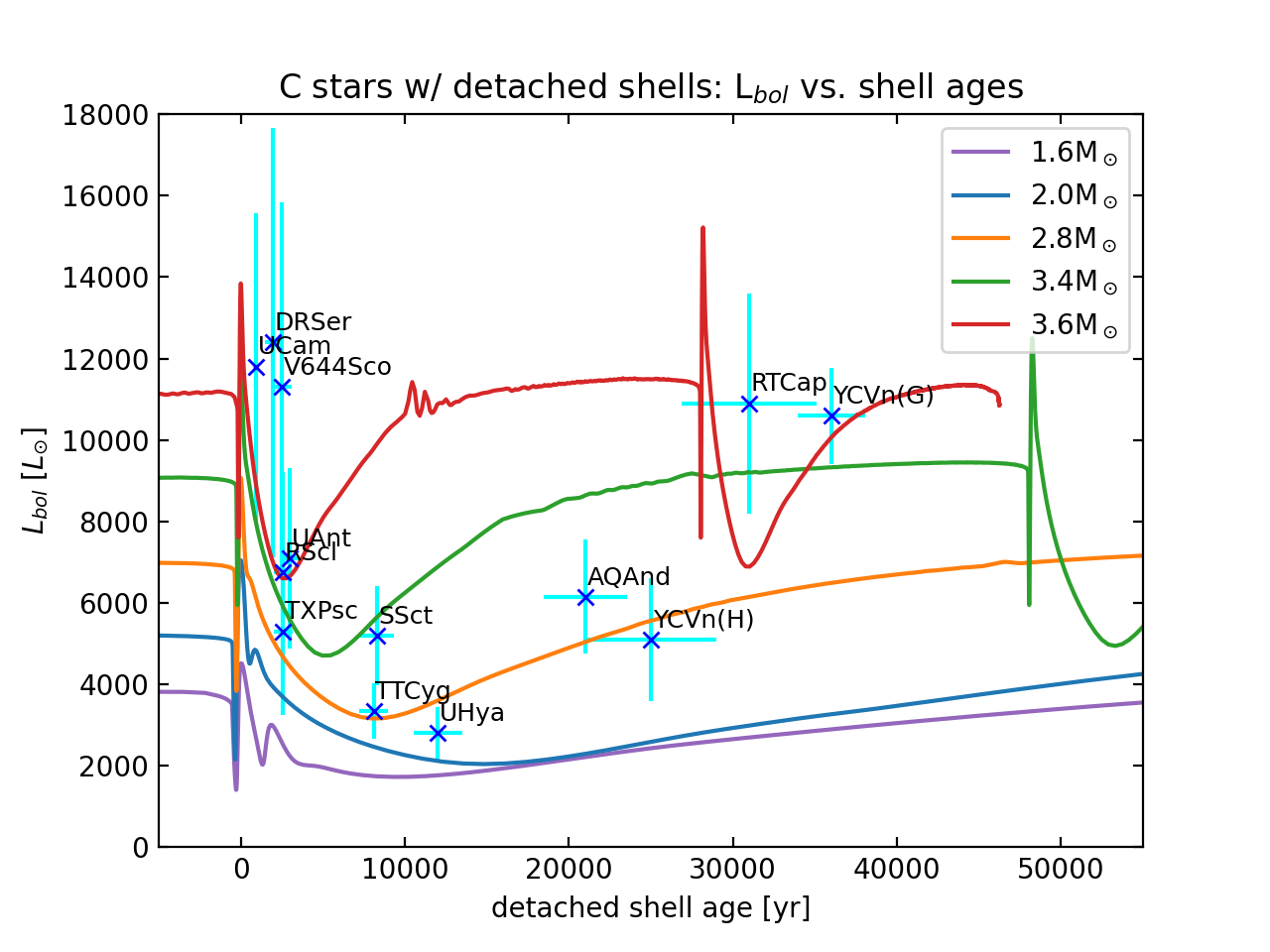}
\end{center}
\caption{Detached shell C star luminosities vs.\ shell dynamical ages, as in Fig.~\ref{fig:LbolVsAge}, here compared with MESA models of the temporal evolution of individual TPs on the upper AGB for progenitor masses of 1.6, 2.0, 2.8, 3.4, and 3.6 $M_\odot$. The model $L_{\rm bol}$ time series are displayed such that the onset of the TPs are aligned at $\tau = 0$. The two points plotted for Y CVn again represent its positions in ($L$, $\tau$) space assuming Hipparcos (H) vs.\ Gaia (G) distances.}
\label{fig:LbolVsAgeModels}
\end{figure}

The late AGB portions of the  2.0, 2.8, and 3.4 M$_\odot$ models are illustrated in Fig.~\ref{fig:MESA}. 
%Mass loss effectively truncates the AGB evolution --- the resulting abrupt cutoffs are evident in Fig.~\ref{fig:MESA} --- \emily{the MESA models were artificially truncated - given more time their luminosities would fall to that of a WD, but their ages would remain virtually unchanged.} and, hence, the duration of the AGB stage in the models is largely determined by the assumed mass loss rate evolution. 
As our focus is on AGB luminosity evolution, we did not attempt to include dredge-up and (hence) the evolution of surface C/O abundance ratio in the modeling; we note that previous approaches to modeling the relationship between TPs and C star shell ejections have assumed surface C/O $>$1, rather than attempting to model the surface C/O abundance evolution  \citep[e.g.,][]{Mattsson2007}. For these simulations, we adopted the standard assumption that the AGB mass loss rate follows a ``Bl\"ocker law'' \citep[i.e., Eq.~16 in][]{Bloecker1995}, with the mass loss rate coefficient fine-tuned to match the initial-final mass relation determined by \citet{Cummings2018} \citep[see][]{Wilson2020}. 
%This assumption is clearly not consistent with the observed episodic mass loss behavior that defines detached shell C stars. 
However, for present purposes --- tracing AGB luminosity evolution during thermal pulses --- the details of mass loss, like surface C abundance, should be unimportant, so long as the internal structure of the star and, in particular, nuclear energy generation processes and rates are accurately modeled. Indeed, 
Fig.~\ref{fig:MESA} confirms that the emergent $L_{\rm bol}$ predicted by these AGB star models are fully consistent with previous work \citep[see, e.g., Fig.~5 in][]{Bloecker1995}. 

Fig.~\ref{fig:MESA} also illustrates several fundamental aspects of theoretically predicted TP AGB behavior. For example, as progenitor mass increases, the overall rise to peak AGB luminosity is steeper (faster), and this peak luminosity increases. Perhaps more importantly, for present purposes, it is immediately apparent from Fig.~\ref{fig:MESA} that once thermal pulsing begins, the characteristic frequency of the TP ``cycles'' is higher (i.e., the TP duration is shorter) for increasing progenitor mass. Also notable is the evolution in the shape of an individual TP luminosity profile for each model: the onset of a TP takes on an increasingly impulsive profile, with sharper (shorter) peaks in luminosity, as the TP portion of AGB evolution proceeds.
This behavior reflects the fact that the convective transport timescale in fully convective AGB envelopes is likely on the order of years \citep{Wilson2019}, such that the energy release from a TP (He shell flash) is essentially instantaneous.

In Fig.~\ref{fig:LbolVsAgeModels}, we again plot $L_{\rm bol}$ vs.\ $\tau_s$ for each detached shell C star (as in Fig.~\ref{fig:LbolVsAge}), but here overlaid with the luminosity vs.\ time profiles of individual thermal pulses, as extracted from the 2.0, 2.8, and 3.4 M$_\odot$ MESA models just described. We also overlay TPs extracted from 1.6 and 3.6 M$_\odot$ MESA models. The $L_{\rm bol}$ time series shown for each model in Fig.~\ref{fig:LbolVsAgeModels} is the most luminous TP that is modeled in its entirety; in the case of the 2.0, 2.8 and 3.4 M$_\odot$ models, these are the penultimate TPs displayed in Fig.~\ref{fig:MESA}. The model $L_{\rm bol}$ time series are displayed such that the onset of these TPs are aligned at $\tau = 0$. It is readily apparent from this Figure that the collective luminosity evolution of detached shell C stars as a function of shell dynamical age is quite consistent with the theoretically predicted temporal profiles (i.e., ``light curves'') of individual thermal pulses for the 2.8 M$_\odot$, 3.4 M$_\odot$, and (in the case of the highest-luminosity stars) 3.6 M$_\odot$ models. Indeed, the models appear to well reproduce specific details of the collective luminosity behavior of detached shell C stars: in particular, an initial peak $L$ approaching or exceeding $\sim$$10^4$ $L_\odot$, followed by a rapid (few thousand year) decline to luminosities of a few $\times$1000 $L_\odot$ and, subsequently, a slow (few $\times$$10^4$ yr) recovery to pre-TP luminosity levels. 

Furthermore, the comparison between predicted and observed $L_{\rm bol}$ vs.\ $\tau_s$ behavior illustrated in Fig.~\ref{fig:LbolVsAgeModels} appears to constrain the range of progenitor masses that characterize detached shell C stars. Specifically, the least luminous known detached shell C stars, TT Cyg and U Hya, lie (respectively) along and just below the 2.8 $M_\odot$ curve; whereas the most luminous stars --- U Cam, DR Ser, V644 Sco, RT Cap, and (adopting its Gaia distance) Y  CVn --- lie along or just above the 3.6 $M_\odot$ curve (though the TP duration for this model appears too short to be consistent with the large shell dynamical age of Y CVn, if one adopts its Gaia distance\footnote{The discrepancy becomes worse if one accounts for the likelihood that the detached shell of Y CVn has undergone significant interactions with the ambient ISM \citep{Libert2007, Matthews2013}.}). Hence, Fig.~\ref{fig:LbolVsAgeModels} indicates that the dozen known examples of detached shell C stars are most likely descended from $\sim$2.5--4.0 $M_\odot$ progenitors.

In this same regard, it is intriguing that the stars S Sct, RT Cap, and U Ant display secondary dust shells with radii that range from roughly twice (S Sct, RT Cap) to $\sim$6 times larger (U Ant) than those of their brighter, inner detached shells \citep{Mecina2014,Izumiura1997}. It is tempting to conclude that these faint, outlying dust shells were ejected by earlier TPs. If so, this might constrain the interpulse timescales of these stars and, hence, their progenitor masses. We note that the TP light curve for an AGB model with progenitor mass of 3.6 $M_\odot$ has a pre-TP luminosity of $\sim$11000 $L_\odot$ and a TP ``period'' of $\sim$2.8$\times10^4$ yr. These parameters closely correspond to the approximate luminosity of RT Cap and the approximate characteristic timescale of its shell ejection sequence (assuming $v_s \approx 8$ km s$^{-1}$). Similarly, the ratio of concentric dust shell radii inferred for U Ant ($\sim$6; Izumuira et al. 1997) is consistent with its position along the 3.6 $M_\odot$ model TP light curve in Figure 4; this progenitor mass lies within the range inferred by Izumuira et al. (1997) from similar (model TP interval) considerations. On the other hand, the present-day luminosity of S Sct appears to fall far below what the models would predict if its interpulse timescale is as short as is implied by its nested shells (i.e., $\sim$8000 yr, assuming the outer shell has an expansion speed similar to that of the inner shell).

%\newpage
\section{Conclusions}\label{sec:Disc}

We have exploited newly available Gaia parallaxes and photometry, along with archival infrared photometry, to obtain refined estimates of the luminosities of all (12) known detached shell carbon stars. We used the results to examine the relationship between these luminosities and the estimated dynamical ages (ejection times) of the detached shells associated with the 12 stars, which range from $\sim$1000 to $\sim$30000 yr.

We find that C stars with the youngest detached shells (dynamical ages $<$3000 yr) have larger luminosities, on average, than those with intermediate-age detached shells (dynamical ages $\sim$8000--12000 yr); many of the former exceed $\sim$$10^4$ $L_\odot$, whereas the latter all have luminosities of a few $\times$$10^3$ $L_\odot$.
This apparent overall decline in carbon star luminosity over a $\sim$$10^4$ yr timescale following shell ejection supports the hypothesis that the shell ejections are triggered by the sharp luminosity increases, and subsequent declines, that are predicted to define AGB thermal pulses \citep{Olofsson1993,Steffen1998,Mattsson2007}. Furthermore, the luminosities of the C stars with the oldest shells (dynamical ages $\sim$20000--30000 yr) are similar to those of the C stars with young shells, suggesting these stars are in the process of a slow recovery to their pre-TP luminosities. 

Overall, this collective temporal behavior appears to closely follow the predicted temporal profiles (i.e., ``light curves'') of individual thermal pulses, as obtained from models of AGB stars that are descended from 2.0, 2.8, and 3.4 $M_\odot$ progenitors (Fig.~\ref{fig:LbolVsAgeModels}). 
We conclude that detached shell C stars effectively trace the luminosity evolution of AGB thermal pulses. Moreover, based on the correspondence between the observed distribution of detached shell C star luminosities vs.\ shell ages and the model TP temporal profiles, the progenitor masses of the majority of C stars known to exhibit detached shells appear to be restricted to the range $\sim$2.5--4.0 $M_\odot$. The same comparison suggests that first three stars in the sequence in Fig.~\ref{fig:LbolVsAgeModels} (U Cam, DR Ser, and V644 Sco) may be descended from somewhat higher-mass progenitors. Thus, most detached shell C stars may be descended from a narrower range of progenitor masses than that estimated for Galactic C stars generally \citep[i.e., $\sim$1.6--4.5~$M_\odot$;][]{Karakas2014,Marigo2020}. 

Given their potential great utility as probes of AGB star luminosity, mass loss, and chemical evolution, it is noteworthy that detached shell C stars also constitute an especially elite class of AGB star. Specifically, there are $\sim$120 bright C stars within $\sim$1 kpc of the Sun, $\sim$70 of which have detectable circumstellar CO emission indicative of mass loss rates $\gtrsim$$10^{-7}$ $M_\odot$ yr$^{-1}$ \citep[][]{Olofsson1993,Schoier2001}. Hence, the known C stars with detached shells comprise $\lesssim$15\% of the mass-losing C star population in the solar neighborhood. It is likely not coincidental that this fraction is is similar to the fraction of time an intermediate-mass star spends on the TP (as opposed to ``early'') portion of the AGB, according to theory \citep{Bloecker1995}. The combination of the short duration of the TP AGB stage, the narrow progenitor mass range leading to AGB stars with the requisite TP behavior to eject a detached shell, and the efficient coupling between radiation pressure and wind momentum that characterizes C-rich dust might also help explain why there have thus far been relatively few unambiguous detections of detached shells associated with O-rich AGB stars \citep[R Hya, RX Lep, and Y Uma;][]{Hashimoto1998,Matthews2013}, despite the fact that O-rich AGB stars far outnumber C stars in the Galaxy \citep{HofnerOlofsson2018}. 

Indeed, given that engulfment of a close binary companion could truncate AGB evolution well before the TP stage commences \citep[e.g.,][]{Wilson2019}, it is actually surprising that detached shell C stars are not an even rarer breed. However, of the dozen known examples, only R Scl displays clear evidence for the impact of binarity on its mass loss history, in the form of spiral structure imprinted on its detached shell \citep{Maercker2016}. Thus, most C stars with detached shells appear to offer examples of ``pure'' single-star evolution to the tip of the AGB, for stars of initial mass $\sim$2.5--3.5 $M_\odot$. This further implies that the relatively rare (distant) examples of unusually luminous, highly dust-enshrouded Galactic C stars that exhibit large ($\sim$15--30 km s$^{-1}$) CO outflow velocities \citep{Kastner1993} may be AGB stars descended from $\sim$2.5--4.5 $M_\odot$ progenitors that we happen to be observing at or just after (within $\sim$1000 yr of) the luminosity peaks of their thermal pulse cycles.

\acknowledgements{J.H.K. wishes to thank Franz Kerschbaum, Sofia Ramstedt, and Matthias Maercker, the lead organizers of ``A Star Has Evolved: a Conference in the Honour of Hans Olofsson'' (held August 2019 in Sm\"{o}gen, Sweden), as well as Hans Olofsson himself, for providing the inspiration for this paper. The authors also thank the anonymous referee, as well as Rens Waters, Hideyuki Izumiura, and Thibaut Le Bertre, for helpful comments that improved the manuscript. E.W.'s research is supported by National Science Foundation grant NSF-2009713 to RIT.}

%\bibliography{DSCstarPaperRefs.bib}{}
%\bibliographystyle{aasjournal}



\end{document}